# A free real-time hourly tilted solar irradiation data Website for Europe

Jonathan Leloux[1,2,*], Luis Narvarte[1,2], Loreto González-Bonilla[1,2]
[1]Instituto de Energía Solar – Universidad Politécnica de Madrid (IES-UPM), Madrid, Spain.
[2]WebPV, Madrid, Spain.
[*]Corresponding author: jonathan.leloux@ies-def.upm.es

ABSTRACT: The engineering of solar power applications, such as photovoltaic energy (PV) or thermal solar energy requires the knowledge of the solar resource available for the solar energy system. This solar resource is generally obtained from datasets, and is either measured by ground-stations, through the use of pyranometers, or by satellites.
The solar irradiation data are generally not free, and their cost can be high, in particular if high temporal resolution is required, such as hourly data. In this work, we present an alternative method to provide free hourly global solar tilted irradiation data for the whole European territory through a web platform. The method that we have developed generates solar irradiation data from a combination of clear-sky simulations and weather conditions data. The results are publicly available for free through Soweda, a Web interface. To our knowledge, this is the first time that hourly solar irradiation data are made available online, in real-time, and for free, to the public. The accuracy of these data is not suitable for applications that require high data accuracy, but can be very useful for other applications that only require a rough estimate of solar irradiation.

## 1 INTRODUCTION

The engineering of solar power applications, such as photovoltaic energy (PV) or thermal solar energy requires the knowledge of the solar resource available for the solar energy system.

The solar resource is generally obtained from datasets, and is either measured by ground-stations, through the use of pyranometers, or by satellites. The solar irradiation data are generally not free, and their cost can be high, in particular if large quantities of data are required at a high temporal resolution, which is the case of the hourly solar irradiation data that are needed for the performance analysis of thousands of Building Integrated PV (BIPV) systems from monitoring.

Pyranometers provide accurate measurements of hourly global horizontal radiation, if the instruments are properly calibrated and maintained, but measurements taken from pyranometers are not available for all the locations, and the accuracy of the data thus depends on where the closest measurement stations are installed. Their measurements are generally not available through automatized services. They are thus generally not available in real-time.

Satellites provide hourly measurements for the whole European continent, with a spatial resolution varying from 1 data/(1x1km$^2$) to 1 data/(50x50km$^2$). They provide solar irradiation data whose accuracy is generally lower than pyranometers, and which is particularly affected by the local cloud cover.

The solar collectors are generally not installed horizontally, but instead in a titled plane. Most solar engineering applications thus require the knowledge of global tilted hourly irradiation, which is deduced from the application of decomposition models to the horizontal irradiation. These models unfortunately convey uncertainties [1]. A growing number of solar applications require the knowledge of real-time solar irradiation at a low cost.

In this work, we present *Soweda* [2], a free real-time hourly tilted solar irradiation data Website for Europe.

The method that we have developed to generate this solar irradiation data makes use of the combination of clear-sky simulations and weather conditions data.

## 2 METHODOLOGY

The solar irradiation data can be obtained from the combination of clear-sky models [3], and simple and free or low cost weather data that indicate the state of the sky, in particular its cloudiness.

The first step is to generate the solar irradiation that would be received by a horizontal surface under clear-sky conditions, i.e. if there were no clouds around the solar disk preventing the direct component of solar radiation from impinging on it.

Several clear-sky models have been proposed in other previous works [4]. They differ among them by their accuracy, their complexity, and the number of input parameters that they require. We have tested out several of these clear-sky models, and we have chosen to use the Perez-Ineichen model [5], because it is both very simple, sufficiently accurate for this kind of application, and also because it requires very little inputs data. The Perez-Ineichen model allows simulating the clear-sky global horizontal irradiance.

The second step is to take into account the real state of the sky, which is carried out through the incorporation of weather data. The state of the sky is provided by some weather conditions providers in the form of *cloud cover octas*, where 0 octas means clear-sky, and 8 octas means completely overcast. In the present case, the weather conditions are obtained from a webservice provided by Wunderground [6], which is, at least partly, publicly available.

Previous works have established a rough relationship between this cloud cover and the cloud transmittance. We have used a model developed by Daryl Myers at NREL because it is simple and effective [7]. This model uses a simple empirical correlation between cloud cover and cloud transmittance, which is shown on figure 1.

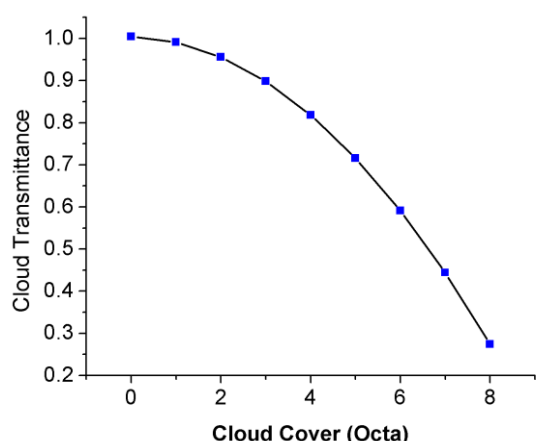

**Figure 1:** Simple empirical relationship between cloud cover and cloud transmittance.

The Global Horizontal Irradiation (*GHI*) is then deduced from the Global Horizontal Irradiation under clear-sky conditions (*GHI$_{CS}$*) and the cloud transmittance (*T$_{cl}$*), as:

$$GHI = GHI_{CS} \cdot T_{cl}$$

The last step is to deduce Global Tilted Irradiation (*GTI*) from *GHI*. This can be done using a decomposition-translation model. We have used the Hay model [8], because it is simple, and it has been reported to perform as well as the most complex models [9].

3 RESULTS

The data are available online from the Graphical User Interface (GUI) of Soweda. The GUI has been created using Wordpress [10] and it uses a responsive layout, which enables it to be used it from any device, including mobile devices with small screens. It is optimized for Google Chrome and Mozilla Firefox.

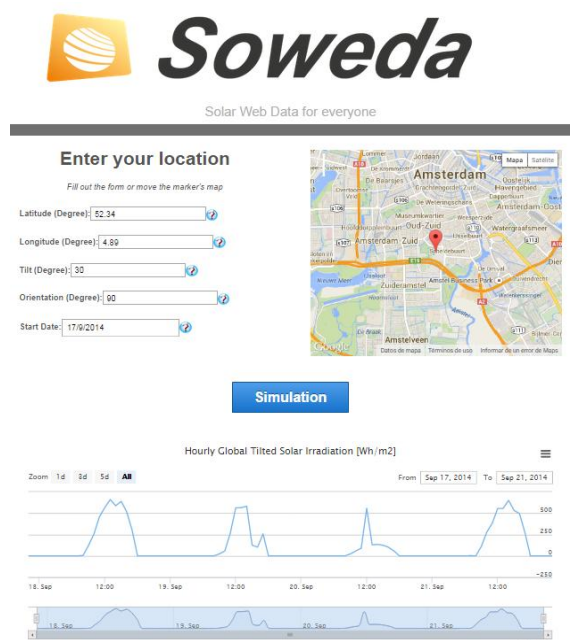

**Figure 2:** User Graphical Interface of Soweda.

The user enters the location through its latitude and longitude, as well as the tilt and orientation angles corresponding to the plane for which she/he needs the solar irradiation data, all expressed in degrees, positive for North and East, and negative for South and West. The location that has been selected is immediately indicated on Google Map. The user has also the possibility to directly define the location from Google Map. The toolbox provides solar irradiation data from up to 5 days before current date, and until one hour before current time.

The user then just needs to click on the *Simulation* button to obtain the results. Once this is done, the solar irradiation data appear on an interactive graph created with the HighChart library [11]. There is a button at the top right side of the chart that allows downloading the data in CSV format, or to generate a picture from the graph.

4 DISCUSSION

The main advantage of Soweda is that it is free.
The methodology that has been employed to generate these data can convey important inaccuracies, and these data are therefore not suitable for applications that require high data accuracy. Although, these data can be very useful in other applications that only require a rough estimate of solar irradiation. This methodology also allows forecasting solar irradiation. This is done by simply using cloud cover forecasts in lieu of cloud cover observations.
Web services are also available from WebPV [12] for professional users who are looking for more reliable and automated data in larger amounts.

5 CONCLUSION

Soweda is a free real-time hourly tilted solar irradiation data Website for Europe.
The method that we have developed to generate this solar irradiation data makes use of the combination of clear-sky simulations and weather conditions data.
The data are available for free from the Web GUI of Soweda.
The methodology that has been employed to generate these data is not suitable for applications that require high data accuracy, but can be very useful in other applications that only require a rough estimate of solar irradiation.

ACKNOWLEDGMENTS

Thanks to Catherine, who carried out a great proofreading job, as usual. We are grateful to Benjamin Wilkin and Jean-Pierre Huart from APERe for their help on the weather data acquisition. Thanks for Rodrigo Moretón, Jesús Robledo and Alberto Luna from WebPV for their help, time and feedback on our work. This work has been partially supported by European Commission within the project PV CROPS [13] (Photovoltaic Cost r€duction, Reliability, Operational performance, Predicition and Simulation) under the 7th Framework Program (Grant Agreement nº 308468).